\documentclass[aps, prl, twocolumn, showpacs, superscriptaddress, amsmath, amssymb]{revtex4-1}
\usepackage[latin1]{inputenc}
\usepackage{graphicx}
\usepackage{dcolumn}
\usepackage{bm}
\usepackage{color}
\usepackage{tabularx}
\newcolumntype{Y}{>{\centering}X}

\usepackage[cdot,textstyle,amssymb]{SIunits}

\usepackage{sistyle}
\SIunitspace{\cdot}
\SIunitdot{\cdot}
\SIdecimalsign{.}

\usepackage[pdfborder={0 0 0}, colorlinks=true,citecolor=blue,linkcolor=black,urlcolor=blue]{hyperref}

\definecolor{orange}{rgb}{1, 0.625, 0.0625}

\begin{document}

\title{Large field-induced gap of Kitaev-Heisenberg paramagnons in \boldmath$\alpha$-RuCl$_{3}$}

\author{Richard Hentrich}
 \email{r.hentrich@ifw-dresden.de}
\affiliation{Leibniz Institute for Solid State and Materials Research, 01069 Dresden, Germany}

\author{Anja U.B. Wolter}
\affiliation{Leibniz Institute for Solid State and Materials Research, 01069 Dresden, Germany}

\author{Xenophon Zotos}
\affiliation{Leibniz Institute for Solid State and Materials Research, 01069 Dresden, Germany}
\affiliation{ITCP and CCQCN, Department of Physics, University of Crete, 71003 Heraklion, Greece}

\author{Wolfram Brenig}
\affiliation{Institute for Theoretical Physics, TU Braunschweig, 38106 Braunschweig, Germany}

\author{Domenic Nowak}
\affiliation{Department of Chemistry and Food Chemistry, TU Dresden, 01062 Dresden, Germany}
\author{Anna Isaeva}
\affiliation{Department of Chemistry and Food Chemistry, TU Dresden, 01062 Dresden, Germany}
\author{Thomas Doert}
\affiliation{Department of Chemistry and Food Chemistry, TU Dresden, 01062 Dresden, Germany}

\author{Arnab Banerjee}
\affiliation{Quantum Condensed Matter Division, Oak Ridge National Laboratory, Oak Ridge, TN, USA}

\author{Paula Lampen-Kelley}
\affiliation{Materials Science and Technology Division, Oak Ridge National Laboratory, Oak Ridge, TN, USA}
\affiliation{Department of Materials Science and Engineering, University of Tennessee, Knoxville, TN, USA}

\author{David G. Mandrus}
\affiliation{Materials Science and Technology Division, Oak Ridge National Laboratory, Oak Ridge, TN, USA}
\affiliation{Department of Materials Science and Engineering, University of Tennessee, Knoxville, TN, USA}

\author{Stephen E. Nagler}
\affiliation{Quantum Condensed Matter Division, Oak Ridge National Laboratory, Oak Ridge, TN, USA}

\author{Jennifer Sears}
\affiliation{Department of Physics and Center for Quantum Materials, University of Toronto, 60 St. George St., Toronto, Ontario, Canada M5S 1A7}

\author{Young-June Kim}
\affiliation{Department of Physics and Center for Quantum Materials, University of Toronto, 60 St. George St., Toronto, Ontario, Canada M5S 1A7}

\author{Bernd B\"uchner }
\affiliation{Leibniz Institute for Solid State and Materials Research, 01069 Dresden, Germany}
\affiliation{Institute of Solid State Physics, TU Dresden, 01069 Dresden, Germany}
\affiliation{Center for Transport and Devices, TU Dresden, 01069 Dresden, Germany}

\author{Christian Hess}\email{c.hess@ifw-dresden.de}
\affiliation{Leibniz Institute for Solid State and Materials Research, 01069 Dresden, Germany}
\affiliation{Center for Transport and Devices, TU Dresden, 01069 Dresden, Germany}

\date{\today}

\begin{abstract}
The honeycomb Kitaev-Heisenberg model is a source of a quantum spin liquid with Majorana fermions and gauge flux excitations as fractional quasiparticles. In the quest of finding a pertinent material, $\alpha$-RuCl$_{3}$ recently emerged as a prime candidate. Here we unveil highly unusual low-temperature heat conductivity $\kappa$ of $\alpha$-RuCl$_{3}$: beyond a magnetic field of $B_c\approx7.5$~T, $\kappa$ increases by about one order of magnitude, resulting in a large magnetic field dependent peak at about 7~K, both for in-plane as well as out-of-plane transport. This clarifies the unusual magnetic field dependence unambiguously to be the result of severe scattering of phonons off putative Kitaev-Heisenberg excitations in combination with a drastic field-induced change of the magnetic excitation spectrum. In particular, an unexpectedly large energy gap arises, which increases approximately linearly with the magnetic field and reaches a remarkably large $\hbar\omega_0/k_B\approx 50$~K at 18~T.
\end{abstract}

\keywords{some keywords}

\maketitle

Topological quantum spin liquids (QSL) are characterised by massive quantum entanglement of states and constitute peculiar states of matter where quantum fluctuations are so strong that even in the ground state a magnetic long-range ordering is suppressed. Amazingly, despite the inherent quantum disorder, the QSL are conjectured to possess well-defined quasiparticles. These are highly non-trivial, because unlike classical systems, the QSLs' quasiparticles arise from the fractionalisation 
in a ground state with topological degeneracy and may have anyonic statistics \cite{Lee2008,Balents2010,Savary2017}. 
Since QSL ground states are experimentally elusive, the detection and rationalisation of just these QSL quasiparticles appear as the natural path towards identifying a QSL system.

Heat conductivity experiments constitute one of the few probes to study such quasiparticle physics because they provide information on the quasiparticles' specific heat, their velocity, and their scattering \cite{Qi2009}\footnote{Energy density is a local operator and will {\it not} probe topological degeneracy, however, fractionalisation will certainly leave characteristic fingerprints in heat transport.}.

In fact, such experiments have been very revealing in clarifying the unconventional ballistic heat-transport characteristics of one-dimensional spinon excitations, the fractional excitations of the spin-1/2 chain \cite{Hess2007,Hlubek2010}, and signatures of unconventional spin heat transport in QSL candidate materials which realise spin-1/2 triangular lattices \cite{Yamashita2009,Yamashita2010} or spin-ice systems \cite{Kolland2012,Toews2013}.

Experimental realisations of QSLs generally are rare. In the quest of finding a pertinent
material to experimentally investigate their physics, $\alpha$-RuCl$_3$ recently emerged as a prime candidate for hosting an approximate Kitaev QSL with Majorana fermions and
gauge flux excitations as  new kinds of fractional quasiparticles \cite{Kitaev2006,Baskaran2007,Knolle2014,Banerjee2016,Nasu2016}.

In this material, strong spin-orbit coupling and an edge-sharing configuration of RuCl$_{6}$ octahedra yield a honeycomb lattice of  $j_\mathrm{eff}=1/2$ states with dominant Kitaev interaction \cite{Plumb2014,Koitzsch2016,Banerjee2016,Yadav2016}. Long-range magnetic order at $T_N\approx7$~K occurs in as-grown samples of $\alpha$-RuCl$_3$ without stacking faults \cite{Sears2015,Kubota2015,Cao2016}, implying a certain degree of further exchange interaction \cite{Banerjee2016,Yadav2016}. Remarkably, a moderate in-plane magnetic field of $\sim8$~T, which is far away from full polarisation  \cite{Johnson2015}, is sufficient to completely suppress the long-range magnetic order \cite{Johnson2015,Majumder2015,Kubota2015,Baek2017}.

In order to probe the emergence of unusual quasiparticles in this putative Kitaev-QSL, we have measured the thermal conductivity $\kappa$ of $\alpha$-RuCl$_{3}$ single crystals in magnetic fields up to $\SI{18}{T}$. 
Overall, four samples (labeled I to IV) from different crystal growth laboratories have been scrutinised. All samples were of high crystalline quality, 
evidenced by the onset of magnetic long range order in the range $T_N=\SI{7.0}{K}$ (sample II) to $T_N=\SI{7.4}{K}$ (sample I), see \textit{Methods} for details.
For fields applied parallel to the planes of $\alpha$-RuCl$_{3}$, we observe a strong impact of the magnetic field on $\kappa$, which upon exceeding $B_c\approx7.5$~T, i.e. in the absence of magnetic order, exhibits a qualitatively new behavior: a low-temperature peak arises in the temperature dependence of $\kappa$ which grows  with magnetic field. The analysis of our data unambiguously implies a radical change in the low-energy spectrum of magnetic excitations, i.e. the opening of an energy gap at $B>B_c$ which increases approximately linearly with the magnetic field.

\section{Results}

\begin{figure}[t]
\centering
\includegraphics[width=\columnwidth]{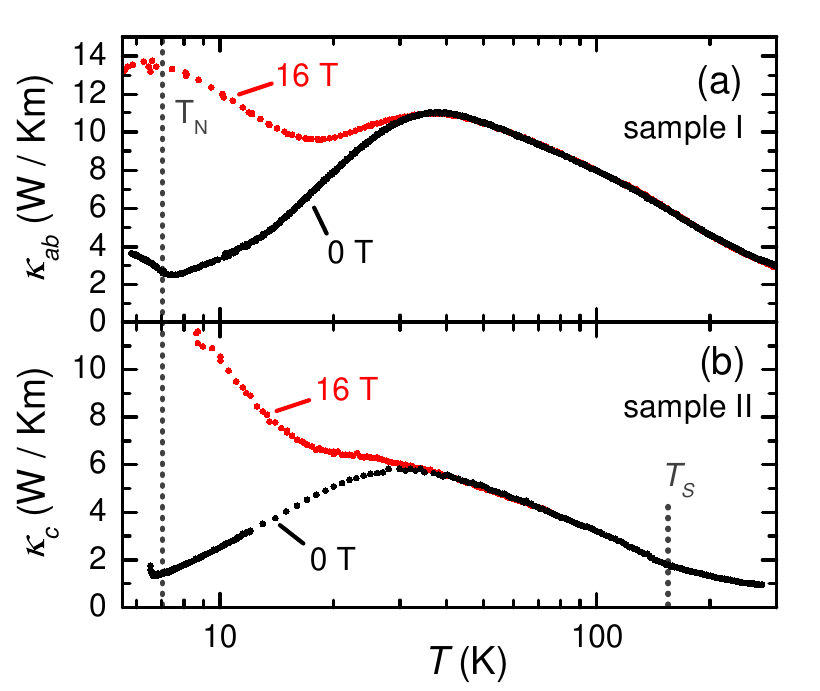}
\caption{\label{fig:zerofield}
Temperature dependence of the heat conductivity of $\alpha$-RuCl$_3$ at zero magnetic field and at $B=\SI{16}{T}$. The heat current was aligned \textbf{(a)} parallel to the $ab$-direction for sample I ($\kappa_{ab}$) and \textbf{(b)} perpendicular to it for sample II  ($\kappa_c$). In both cases, the field was applied parallel to the $ab$-planes and perpendicular to the heat current. The onset of long range magnetic order at $T_N\approx7$~K and the structural transition at $T_s\approx155$~K are indicated.}
\end{figure}

The upper panel of Figure \ref{fig:zerofield} shows representative data of the in-plane thermal conductivity $\kappa_{ab}$ of $\alpha$-RuCl$_{3}$ as a function of temperature $T$ in zero field (see the supplementary information (SI) for $\kappa_{ab}$ of other single crystals with essentially the same temperature dependence) and at $B=16$~T, applied parallel to the $ab$-planes. In zero magnetic field, upon cooling from $300$~K down to the base temperature ($5.5$~K) of our setup, $\kappa_{ab}$ increases steadily up to a distinct maximum at around $\SI{40}{K}$, and decreases steeply at lower temperature. A kink around $7.5$~K coincides roughly with the onset of long-range magnetic order of the system \cite{Cao2016}.

At first glance, these zero magnetic field data of $\kappa_{ab}(T)$ at $T>T_N$ with a single peak structure resemble that of a conventional phononic heat conductor \cite{Berman}: In such a case the phononic heat conductivity, which can coarsely be estimated as $\kappa_\mathrm{ph}\sim c _Vvl$, increases strongly with temperature in the low-$T$ regime, where the phononic velocity $v$ and mean free path $l$ are essentially temperature independent, with the phononic specific heat $c_V$. Towards higher $T$, phonon umklapp processes increasingly limit $l$, resulting in a broad peak in $\kappa_\mathrm{ph}$ followed by a fast decline. For antiferromagnetic insulators it is well known that scattering of phonons off paramagnon fluctuations of the incipient long range order may give rise to a significant suppression of $\kappa_\mathrm{ph}$ above and a recovery below the N\'{e}el ordering temperature, respectively \cite{Slack58,Laurence73,Slack61,Steckel2014,Hess99}. The whole $\kappa_{ab}(T)$ including the observed kink at $T_N$ seems perfectly in line with such a scenario. 

Strikingly, the application of a large in-plane magnetic field of $B=16$~T, at which magnetic order is absent, dramatically changes $\kappa_{ab}$, and thereby challenges such a rather conventional interpretation:
$\kappa_{ab}$ is drastically enhanced at low temperature - a second large peak emerges at around $7$~K which even exceeds the one at higher temperature.

\begin{figure}[t]
\centering
\includegraphics[width=0.8\columnwidth]{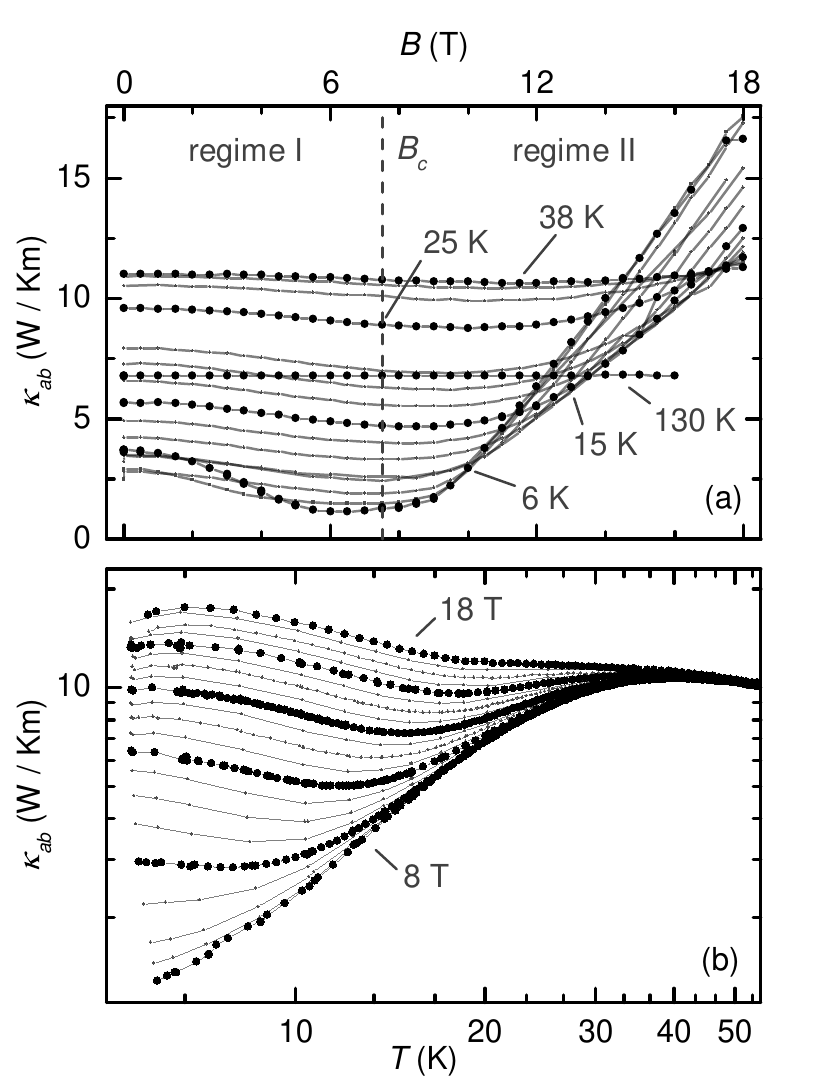}
\caption{Field and temperature dependence of the heat conductivity of $\alpha$-RuCl$_3$ (sample I) with the heat current and the magnetic field $B$ parallel to the $ab$-direction. (a) Isothermal field dependent heat conductivity $\kappa_{ab}(B)$ as a function of magnetic field for selected temperatures. 
 (b) Temperature dependence $\kappa_{ab}(T)$ for $B > 7.5$~T, measured at $B$-steps of 0.5~T. 
\label{fig:kappa_all}
}
\end{figure}

The unconventional nature of the field-induced double-peak structure in the temperature dependence of $\kappa_{ab}$ is further confirmed by a detailed mapping of $\kappa_{ab}(T,B)$ up to an in-plane field of $B=18$~T, as is shown in Fig.~\ref{fig:kappa_all}. Apparently, the impact of the magnetic field on the heat transport is profoundly different for $B\lesssim 7.5$~T on the one hand and $B\gtrsim 7.5$~T on the other hand, which clearly defines two field regimes (labeled I and II, respectively) which are separated by a critical field $B_c\approx7.5$~T. In regime I, as is evident from panel (a) of Fig.~\ref{fig:kappa_all}, $\kappa_{ab}$ slightly decreases for all temperatures $T<\SI{40}{K}$ upon increasing the field from zero to $B_c$. This suppression is most pronounced at $T=6$~K where it reflects the suppression of the long range magnetic order. A dramatically different field dependence occurs upon further increasing the field (regime II), where $\kappa_{ab}$ strongly increases with increasing field. Remarkably, for $T\lesssim 15$~K, this increase is essentially linear in magnetic field up to 18~T.

\section{Magnetic heat transport at high magnetic field?}

Without any doubt, the most prominent feature of the present data is the large field-induced low-temperature peak in $\kappa_{ab}$ in regime II, which increasingly grows with field (Fig.~\ref{fig:kappa_all}b). Note that the critical field $B_c$, which marks the onset of this regime, coincides with  the complete field-induced suppression of long-rang magnetic order which governs the lowest temperature physics in regime I but is absent in regime II \cite{Johnson2015,Majumder2015,Kubota2015,Baek2017}. Thus, the low-temperature peak in regime II must be of a qualitatively different origin in contrast to the low-temperature upturn in regime I below $T_N$, which is closely related to spin fluctuations in the system.

A priori, two very different scenarios can be invoked for explaining the nature of a double-peak structure in $\kappa(T)$ of an electrical insulator which hosts a fluctuating spin system. On the one hand, this could be the signature of magnetic heat transport that, in turn, leads to a pertinent contribution to the (otherwise conventional phononic) heat conductivity. Such a mechanism is common in low-dimensional systems such as spin chains, ladders, and planes \cite{Sologubenko00,Hess01,Hess03,Hess2007,Hlubek2010,Steckel2016}. In these cases, phonons and magnetic excitations yield two independent transport channels. On the other hand, a double-peak structure is also known to occur in purely phononic heat transport, resulting from the heat carrying phonons scattering off another degree of freedom, such as a spin excitation with a well defined excitation energy $\hbar\omega_0$ \cite{Hofmann2001,Jeon2016}. Such scattering affects the phononic heat transport over a large temperature range, but has its strongest impact in the temperature regime where the energy of the majority of heat carrying phonons coincides with $\hbar\omega_0$.

An unambiguous signature of low-dimensional magnetic heat transport is its anisotropy: spin-heat is transported only along a certain crystal direction along which a significant energy dispersion of the spin excitations exists, i.e., along the directions without a significant magnetic exchange interaction the magnetic heat transport is absent \cite{Sologubenko00,Hess01,Hess03,Hess2007,Hlubek2010,Steckel2016}. We therefore investigated the heat conductivity of $\alpha$-RuCl$_3$ perpendicular to the planes ($\kappa_c$), where the magnetic exchange interaction is negligible (sample II, see Fig.~\ref{fig:zerofield}b). Remarkably, apart from minor differences in details (see Fig.~\ref{fig:caxis} in the SI), we observe practically the same temperature and magnetic field dependence as for $\kappa_{ab}$. The most important finding  is the direct comparability of $\kappa(c)$ to $\kappa(ab)$ at $B=0$ and $16$~T, with the out-of-plane thermal transport exhibiting the same low-temperature enhancement in both cases, i.e. the presence of a new low-temperature peak. Hence, we can exclude the scenario that transport by the emergent elementary excitations of the spin system gives rise to the low-temperature peak in regime II. This means unambiguously that the field-induced low-temperature peak of $\kappa_{ab}$ is primarily phononic, and its peculiar temperature and magnetic field dependence arises from an unusual, field dependent scattering process of the phonons.

\section{Magnetic phonon scattering}
\begin{figure}[t]
\centering
\includegraphics[width=\columnwidth]{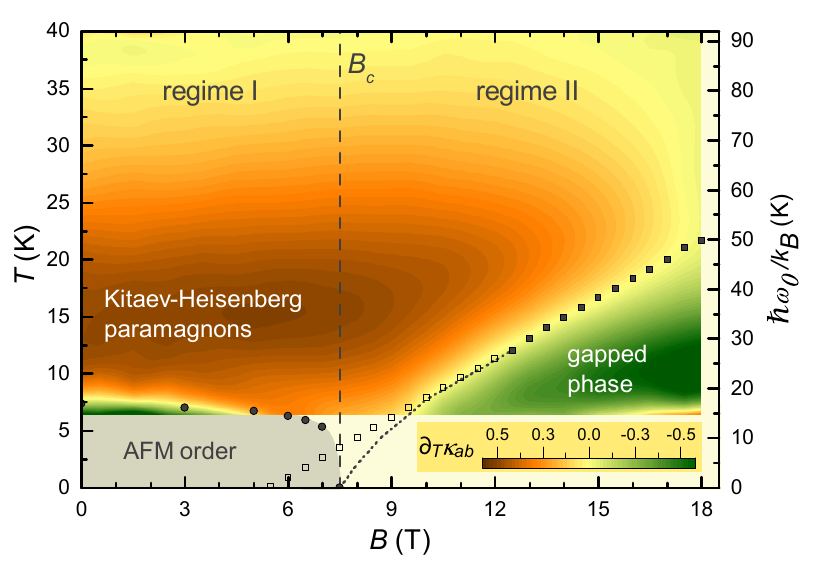}
\caption{False-colour representation of the temperature derivative $\partial\kappa_{ab}(T, B)/\partial T$ (sample I) together with the gap energy $\hbar\omega_0 /k_B$ (squares) as extracted from the phononic fit. The left ordinate shows the temperature $T$ of the measurement, while the right ordinate shows $\hbar \omega _0 / k_B$. Note, that for $B\leq12$~T an unambiguous extraction of $\omega_0$ cannot be obtained from fitting $\kappa_{ab}(T)$ because the $T_\mathrm{min}$ is too close to the lower limit of the measurement range. Nevertheless, good fits to the data are obtained if the $\omega_0(B)$ for $B>12$~T is extrapolated towards smaller fields (open squares) and subsequently used to fit $\kappa_{ab}(T)$ at the corresponding fields (see \textit{Methods} for details). Note that these extrapolated $\omega_0$ should be regarded as an upper limit only. The experimental data can similarly well be described with somewhat smaller $\omega_0(B)$, the field dependence of which is indicated by the dotted line.
\label{fig:gap}
}
\end{figure}

After having established this first important result, we now move on to rationalising the field and temperature dependence of the heat conductivity more thoroughly. 
Without further analysis
and by invoking the above-mentioned magnetic scattering scenario of phonons
one can conclude from the presence of the double-peak structure in regime II with a clear minimum at $T_\mathrm{min}$, 
that an energetically well-defined magnetic mode with energy $\hbar\omega_0$ exists which scatters primarily the heat carrying phonons of that energy.
This can be understood from the fact that the energy of the phonons which predominantly carry heat, $\hbar\tilde{\omega}$, is strongly temperature dependent \cite{Berman}. I.e. the two peaks at lower and higher temperature correspond to $\tilde{\omega}<\omega_0$ and  $\tilde{\omega}>\omega_0$, whereas  $\tilde{\omega}\approx\omega_0$ at $T_\mathrm{min}$.
This immediately suggests that a rough quantitative estimate of the scattering magnetic mode energy $\hbar\omega_0$ can be obtained by reading $T_\mathrm{min}$ off the data.
We therefore plot the temperature derivative $\partial\kappa_{ab}/\partial T$ in false colour representation (Fig.~\ref{fig:gap}). At $B\gtrsim11$~T, the minimum position $T_\mathrm{min}$ depends about linearly upon $B$. At smaller fields, however, $T_\mathrm{min}(B)$ attains a steeper slope and rapidly moves out of the measured temperature window, suggestive of an approximate extrapolation towards $B_c$ at zero temperature (dotted line Fig.~\ref{fig:gap}).
Thus, while the dominant magnetic scattering mode energy $\hbar\omega_0$ seems to be very close to zero around $B_c$ and at smaller fields, it rapidly develops a substantial size at higher magnetic fields.

\begin{figure}[h!]
\centering
\includegraphics[width=\columnwidth]{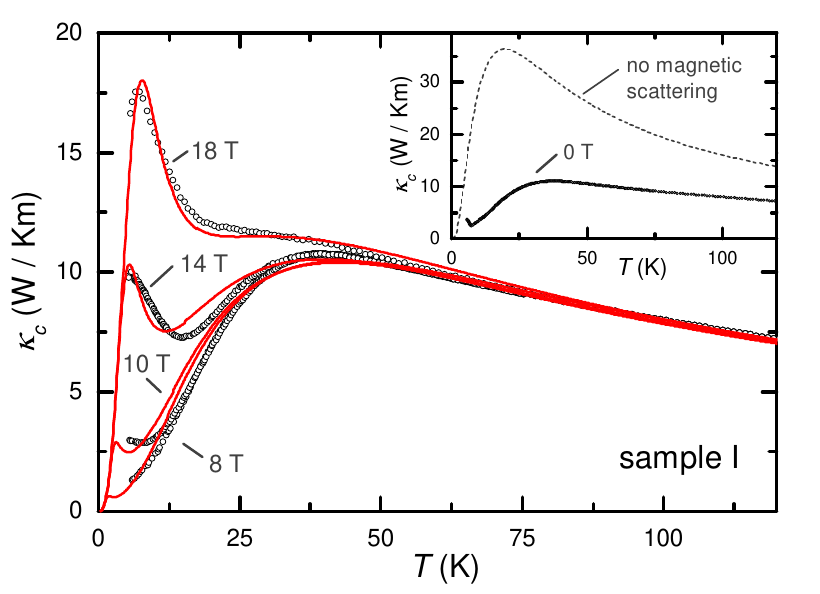}
\caption{$\kappa_{ab}$ data of sample I (open symbols) and fits to the Callaway model (red solid lines) for selected representative magnetic fields. The fits have been obtained by incorporating magnetic scattering of the phonons (see \textit{Methods}). Inset: $\kappa_{ab}$ data of sample I (symbols) at zero field as compared to the hypothetical phononic heat conductivity described by the model if the magnetic scattering mechanism is switched off (dashed line). This highlights that the magnetic scattering affects the phonon heat conductivity in a very large temperature range where the field induced changes occur essentially at $T<50$~K.}
\label{fig:abaxisfits}
\end{figure}

One can exploit $T_\mathrm{min}(B)$ further and estimate the field dependence of the magnetic mode energy $\hbar\omega_0(B)$ quantitatively by considering that for a conventional isotropic phononic system, the majority of heat carrying phonons at a certain temperature $T$ possess an energy of about $ \alpha k_BT$ with $\alpha\approx4$ \cite{Berman}. 
For more anisotropic phononic systems, as one might expect for $\alpha$-RuCl $_3$, simple dimensional considerations (see \textit{Methods}) suggest a somewhat reduced $\alpha$ (in particular, $\alpha\approx2.6$ for a hypothetical purely two-dimensional phononic system). 
Thus, by translating $T_\mathrm{min}(B)$ into the phonon energy which is affected strongest by the magnetic scattering one can directly extract the field dependence of the magnetic mode energy as $\hbar\omega_0(B)\approx\alpha T_\mathrm{min}(B)$, with $\alpha$ roughly in the range of 2.6 to 4. In fact, the low-temperature specific heat of $\alpha$-RuCl$_3$ has been reported to follow a $T^2$ rather than a $T^3$ dependence \cite{Cao2016}, which indicates a significant anisotropy of the phonons. Yet, the anisotropy of the heat transport is only moderate (see Fig.~\ref{fig:zerofield}) as compared to prototype quasi two-dimensional phononic systems such as graphite \cite{Ho1972}, implying a significant interlayer lattice coupling and consequently only a moderate anisotropy for the phononic system, in contrast to the two-dimensional nature of the magnetic system.

A further corroboration of the scenario of magnetic phonon scattering can be obtained by analysing $\kappa(T,B)$ in terms of a phononic model which takes the magnetic scattering into account. We follow the usual approach of Callaway \cite{Callaway61,Callaway59} and express the heat conductivity in terms of an energy dependent combined relaxation time $\tau_c$ which takes various scattering processes into account (see \textit{Methods}). For conventional phononic systems good descriptions of the temperature dependence of $\kappa$ can be achieved if the  standard expressions for the phonon relaxation times describing umklapp, point defect, and boundary scattering are comprised in $\tau_c$.

As one can already conjecture in view of the unconventional double-peak structure of $\kappa_{ab}(T,B)$ at $B>B_c$, a fit of the Callaway model to our data fails in this standard phononic picture. However, a qualitatively reasonable fit is indeed possible if we adapt the model to the afore sketched magnetic scattering scenario by introducing an additional relaxation time $\tau_\mathrm{mag}$ which describes such scattering. More specifically, we assume the phonons to scatter within an energetically broad magnetic excitation spectrum (mimicking the theoretically predicted \cite{Knolle2014,Briffa2016} and experimentally observed \cite{Banerjee2016a} character of the excitation spectrum of the Kitaev model and of $\alpha$-RuCl$_3$, respectively) from a reservoir which is dominated by an energetically sharp and magnetic field dependent low-energy mode at $\hbar\omega_0$ (see \textit{Methods} for details). 

Despite the simplicity of this model, it is indeed possible to simultaneously fit the data in regime II with a field independent parameter set for the usual phonon scattering terms and a field dependent magnetic scattering term $\tau_\mathrm{mag}(B)$, see Fig.~\ref{fig:abaxisfits}.
The values for $\omega_0(B)$ extracted thereby are plotted in Fig.~\ref{fig:gap} as square symbols. Obviously, they display a field dependence similar to that of the minimum in the heat conductivity, $T_\mathrm{min}(B)$, and are in the same energy range as expected from the above considerations with respect to $T_\mathrm{min}$.
We emphasise that a qualitatively similar result is reached upon analysing $\kappa_c$ with the same procedure, which further corroborates this analysis (see SI).

At magnetic fields smaller than $B_c$, the heat conductivity of $\alpha$-RuCl$_3$ is always significantly smaller than in the high-field phase. This straightforwardly implies that in regime I the phonon scattering off the magnetic excitation spectrum is even stronger than in regime II (see Fig.~\ref{fig:zerofield}). 
In view of the vanishing $\omega_0$ at around $B_c$, this suggests that in regime I the scattering magnetic modes are at relatively low energy if not gapless. Evidently, the 
field induced changes of $\kappa_{ab}$ in this phase are relatively small (see Fig.~\ref{fig:kappa_all}). This small field dependence then naturally is consistent with field induced changes of the magnetic spectrum, connected with the gradual suppression of long range magnetic order. These are more subtle than those apparently induced by high magnetic fields.

\section{Conclusion}
In conclusion, our data and the subsequent analysis clearly show that the heat transport of $\alpha$-RuCl$_3$ is primarily of phononic type. More specifically, the field-induced low-temperature peak in the heat conductivity cannot be explained by the expected 
exotic excitations of a putative Kitaev-Heisenberg QSL carrying heat. 
Nevertheless, the magnetic excitations of $\alpha$-RuCl$_3$ dramatically impact the phononic heat transport along all directions through scattering of the phonons off the magnetic excitation spectrum. 
This scattering is particularly strong in regime I,
which at first glance seems to be consistent with the incipient long range magnetic order. However, the magnitude of the low-temperature increase of $\kappa$ in the ordered phase is relatively small as compared to the dramatic enhancement in regime II. This implies that even in the magnetically ordered phase considerable magnetic degrees of freedom exist which is in line with the significantly reduced magnetic moment observed in inelastic neutron scattering  \cite{Banerjee2016}. These residual degrees of freedom scatter the phonons and are likely to remain quantum disordered all the way down to zero energy. 
Since the phonon heat conductivity at low temperature primarily is carried by acoustic phonons with small momenta $k\sim0$, it seems natural to conclude that the low energy paramagnons relevant for the scattering possess small momenta as well.
On the other hand, the dramatic enhancement of the heat conductivity at higher fields in regime II ($B>B_c$) implies that these low-energy excitations are increasingly gapped out, i.e. the strongest field-induced change of the excitation spectrum concerns the excitations close to the $\Gamma$-point.
The field-induced spin gap is surprisingly large, since at 18~T it is already of the order of the extracted Kitaev interaction of $\alpha$-RuCl$_3$ \cite{Banerjee2016}. One might thus speculate that the field-induced phase at $B>B_c$ is  governed by new physics where the emergent quasiparticles are indeed different from those of the Kitaev-Heisenberg-paramagnons at $B<B_c$.

Final note: 
Upon finalizing this manuscript, we became aware of another study of the field dependence of the heat conductivity by Leahy et al. \cite{Leahy2016} where a similar enhancement of $\kappa_{ab}$ in the field regime II is reported and is interpreted as the signature of in-plane heat transport by massless so-called \textit{proximate Kitaev spin excitations}.
We stress that the latter interpretation for the field-induced low-temperature peak can be ruled out since this peak is essentially isotropic in $\kappa_{ab}$ and $\kappa_c$ and therefore clearly of phononic origin, as is explained above in detail.

\section{Methods}

\subsection{Sample preparation}
Samples I and III have been grown at the TU Dresden. For the synthesis, pure ruthenium-metal powder ($\SI{99.98}{\%}$, Alfa Aesar) was filled into a quartz ampoule under argon atmosphere, together with a sealed silica capillary containing chlorine gas ($\SI{99.5}{\%}$ Riedel\text{-}de Ha\"{e}n). The molar ratio of the starting materials was chosen to ensure in-situ formation of RuCl$_{3}$ and its consequent chemical transport according to the reaction: RuCl$_{3}$(s) $+$ Cl$_{2}$(g)$\rightarrow$ RuCl$_{4}$(g) \cite{Binnewies2012}. The vacuum-sealed reaction ampoule ($p \approx \SI{0.1}{Pa}$) was subsequently shaken in order to break the chlorine-containing capillary and to release the gas. The ampoule was kept in the temperature gradient between $T_1 = \SI{750}{^\circ C}$ (starting material) and $T_2 = \SI{650}{^\circ C}$ for 5 days. The value of $T_2$ was optimised to rule out transformation of the product into the $\beta$-modification. The value of $T_1$ was chosen such as to avoid decomposition of RuCl$_{3}$ into elemental ruthenium and RuCl$_{4}$ gas species \cite{Oppermann2005}. The crystalline product represented pure $\alpha$-RuCl$_{3}$ (according to powder X-ray diffraction) without inclusions of ruthenium. Single-crystal X-ray diffraction studies (Apex II diffractometer, Bruker-AXS, Mo K$\alpha$-radiation) and EDXS (Oxford Silicon drift detector X-MaxN, Hitachi SU 8020 SEM, $\SI{20}{kV}$) of the crystals have confirmed a monoclinic structure \cite{Johnson2015} and the nominal composition.
The crystals are black with shiny surfaces, of millimeter size along the ab-plane, and of a thickness less than 0.1 mm.
These crystals have been cut into rectangular shape as to have the optimum experimental geometry.

Sample II has been grown at Oak Ridge National Laboratory \cite{Banerjee2016}. It is significantly larger (roughly $\SI{13x1.8x2}{mm^3}$) and thus well suited for studying $\kappa _c$. For the measurement the crystal has been mounted as-grown.

Sample IV has been grown at the University of Toronto \cite{Sears2017}. It has similar dimensions to samples I and III, roughly a rectangular shape and therefore has been measured as-grown. 

The crystal quality of all sample batches has been checked rigorously by means of x-ray diffraction and susceptibility measurements (not shown here). All sample handling has been done with great care to avoid the introduction of crystal defects such as stacking faults, which can reportedly be induced by putting  $\alpha$-RuCl$_{3}$ single crystals under mechanical strain \cite{Cao2016}.

\subsection{Determination of the N\'{e}el temperature}
In order to confirm the connection of the minimum in $\kappa_{ab}(T)$ for $B\leq7.5$~T with the N\'{e}el temperature $T_N$ of $\alpha$-RuCl$_{3}$, we have measured the temperature dependence of the magnetic susceptibility and of the specific heat on a sample of the same batch as samples I and III as a function of magnetic field (applied parallel to the $ab$-planes).
The magnetic susceptibility has been obtained for a single crystal ($m \approx SI{4}{mg}$) of the same batch as samples I and III in external magnetic fields $B=\mu_0 H = \SI{0}{T}$-$\SI{7}{T}||ab$ using a Superconducting Quantum Interference Device-Vibrating Sample Magnetometer (SQUID-VSM) from Quantum Design. The extracted values for $T_N = d (\chi\times T)/dT$ are in good agreement with the specific heat results.

The specific heat measurements have been performed on a piece of the same single crystal used for the magnetic susceptibility studies ($m\approx \SI{2.5}{mg}$) between 1.9~K and 20~K using a heat-pulse relaxation method in a commercial Physical Properties Measurement System (PPMS) from Quantum Design. The magnetic field $\mu_0 H = \SI{0}{T}$-$\SI{9}{T}$ has been applied in the $ab$ direction, for which an additional sapphire block has been used for mounting the sample. The heat capacity of the complete sample holder (addenda) had been determined prior to the measurements for the purpose of separating the heat capacity contribution of the sample from the total heat capacity of the setup.
Note that the cutting of the sample induced a small amount of stacking faults in the single crystal, which, however, can be well-separated from the intrinsic signal around 7.4 K due to its shifted $T_N$ to slightly higher temperatures. At an external magnetic field of 7.5 T, no signs of a magnetic transition can be detected in our sample in the full temperature regime. Due to a broadening of the magnetic transition in higher fields, we restrain from an entropy-conservation construction to obtain $T_N$. Instead, the transition temperature is determined from the maximum position of the peak in the specific heat capacity. The difference between both constructions is less than 0.2 K for $T_N$ in zero field.

For clarity, only the results of the specific heat are plotted in Fig.\ref{fig:gap}, which at $B=7.5$~T confirm the absence of a thermodynamic transition down to 1.9~K.

\subsection{Heat conductivity measurements}

Steady state thermal conductivity measurements have been performed in the temperature range $\SI{5.5}{K}$-$\SI{300}{K}$, utilising a home made vacuum setup optimised for low noise measurements in a standard four point probe geometry \cite{Hess03a}.
The in-plane thermal conductivity $\kappa _{ab}$ has been measured for samples I, III and IV, the out-of-plane component $\kappa _{c}$ for sample II. Magnetic fields up to $\SI{18}{T}$ have been applied parallel to the honeycomb planes for sample I and II and perpendicular for sample III using a standard $^4$He bath cryostat with a superconducting magnet coil. The delicateness of the samples requires special means to obtain high quality data. All samples thus have been cooled/heated at slow rates to minimise thermal stress. Furthermore, data have been recorded during heating as well as cooling the samples to rule out hysteretic effects. 
All these measures underpin the robustness of our results which is evident by the high reproducibility we observe in our $\kappa _{ab}$-data. Temperature sweeps at constant field as well as magnetic field sweeps at constant temperature have been undertaken, resulting in a large, consistent data set.

\subsection{Modelling of phonon heat conductivity}

The phonon heat conductivity $\kappa_\mathrm{ph}$ of a three-dimensional isotropic solid can be described by \cite{Ziman}:

\begin{equation}
\kappa_\mathrm{ph} =  \frac{1}{3 (2 \pi)^3} \ \int c_{\bm k} v_{\bm k} l_{\bm k} \ d\bm{k}
\end{equation}
where $c_{\bm k}=\frac{d}{dT}u_{\bm k}$, $v_{\bm k}$, and $l_{\bm k}=v_{\bm k}\tau_{\bm k}$ denote the contribution to the specific heat, the velocity, and the mean free path of a phononic mode with wave vector ${\bm k}$. $u_{\bm k}$ and $\tau_{\bm k}$ are the mode's energy and relaxation time, respectively. 

The Callaway Model \cite{Callaway59,Callaway61} refers to the evaluation of the above by applying a Debye ansatz for the phonon heat capacity, assuming $v_{\bm k}  =v_s $ (the speed of sound) for all phonon branches, and by introducing an energy dependent relaxation time $\tau_c$. This yields the low $T$ approximation

\begin{equation}
 \kappa  (T) = \frac{k_B}{2\pi^2v_s}\left(\frac{k_BT}{\hbar}\right)^3 \int_0^{\Theta_D/T}
\frac{x^4e^x}{\left(e^x-1\right)^2}\tau_c(x)\ dx, \label{Callaway}
\end{equation}

with  Boltzmann's constant $k_B$, Planck's constant $\hbar$, and $x = \hbar\omega / k_BT$.

Before further specifying the effective phonon scattering rate $\tau ^{-1} _c$, a general statement can be deduced from the analytical form of the first term of the integrand. It has a distinct peak at $\hbar\omega \approx 4 k_B T$, determining roughly the energy of those phonons which contribute most to the heat transport at a given temperature $T$ \cite{Berman}. Thus, in presence of a magnetic scatterer with well defined energy $\omega _0$, such scattering can be expected to have the strongest impact on the heat transport at around $T\approx\hbar\omega_0/(4k_B)$.

The layered crystal structure of $\alpha$-RuCl$_{3}$, with its strong intralayer and weak van der Waals interlayer coupling suggests an anisotropic phononic structure with a significantly lower phonon velocity perpendicular to the planes. For the hypothetical extreme case of just two-dimensional phonon propagation, a modification of the first part of the integrand in Eq.\ref{Callaway} towards $(x^3e^x)/(e^x-1)^2 $ should be considered. In this case energy of the predominantly heat carrying phonons reduces with respect to the three-dimensional case to $\hbar\omega \approx 2.6 k_B T$. Thus, for a very anisotropic lattice such as $\alpha$-RuCl$_{3}$, the scaling factor $\alpha$ which relates the position of the minimum of $\kappa(T)$ with the energy of a magnetic scatterer can be expected to roughly be in the range 2.6 to 4 times $k_BT_\mathrm{min}$.

For conventional phononic systems (i.e. non-magnetic, electrically insulating crystals), the effective phonon scattering time in Eq.\ref{Callaway} is composed of conventional scattering mechanisms, for which empirical expressions are well established, viz. phonon-phonon umklapp scattering $\tau_P^{-1}= A\, T \omega ^3 e^{- \Theta _D/b T}$, phonon-defect scattering $\tau_D^{-1}= D \omega ^4$ and phonon-boundary scattering $\tau_B^{-1}=v_s L^{-1}$. Following Matthiessen's rule the combined effective scattering rate yields as

\begin{equation}
\tau_{c,0}^{-1}=\tau_P^{-1}+\tau_D^{-1}+\tau_B^{-1}. \label{matthiessen}
\end{equation}

As described in the main text, such a conventional model is unable to reproduce the double peak structure of $\kappa _{ab}$ which we observe at high magnetic fields.

\subsection{Modelling of magnetic phonon scattering} 
In order to incorporate the phenomenologically sketched magnetic scattering into the model, we add a further scattering rate  $\tau_\mathrm{mag}^{-1}$ which describes the magnetic scattering of the phonons, i.e.
\begin{equation}
\tau_c^{-1}= \tau_{c,0}^{-1}+\tau_\mathrm{mag}^{-1} \label{comb}
\end{equation}

In lack of a suitable microscopic theory for describing the magnetic scattering processes of phonons in $\alpha$-RuCl$_{3}$, we use an empirical approach to analytically describe $\tau_\mathrm{mag}$, which we adapt from earlier successful modellings of magnetic resonant scattering of phonons off magnetic triplet excitations which are well defined in energy and momentum \cite{Hofmann2001,Jeon2016}:
\begin{equation}
 \tau_\mathrm{mag}^{-1}= C\frac{\omega ^4}{(\omega ^2 - \omega ^2 _0)^2}\frac{e^\frac{-\hbar\omega_0}{k_BT}}{1+3e^\frac{-\hbar\omega_0}{k_BT}}\label{res_triplet},
\end{equation}

with a coupling constant $C$. The second factor $\omega ^4/(\omega ^2 - \omega ^2 _0)^2$ represents a resonant scattering cross section and the last factor expressing the thermal population of the triplet mode, where we neglect Zeeman splitting.
Indeed, a qualitative fit of the peculiar temperature and field dependence of our $\kappa$ data based on the Callaway model (Eq.~\ref{Callaway}) using the combined relaxation rate $\tau_c^{-1}$ as given by Eqs.~\ref{comb} and \ref{res_triplet} is possible, see Fig.~\ref{fig:comp_fit} in the supplementary information. The use of expression for $\tau_\mathrm{mag}$ apparently has, however, several caveats.
Firstly, the chosen resonant scattering cross section describes phonon scattering off an energetically sharply defined magnetic mode at $\omega_0$. However, in quantum magnets with fractional quasiparticles, one typically expects energy-momentum continua rather than a comparatively well-defined energy-momentum dispersions. 
In principle, the weight of the continuum and its boundaries could be $q$-dependent.
We discard this and replace the sharp resonant cross section by a broad function which mimics the excitation spectrum of $\alpha$-RuCl$_{3}$ and of the Kitaev-Heisenberg model with a high-energy cut-off \cite{Banerjee2016a,Knolle2014,Briffa2016}. We therefore replace the resonant factor by the Heaviside function $\theta(K-\hbar\omega)$, with $K$ the high-energy cut-off.

Secondly, one might argue that in view of the excitation continuum 
an integration over frequencies with respect to the occupation number, respecting, however, a low-energy cut-off of $\omega\geq\omega_0$ should be considered. Since the exact nature of the excitation is not known we discard such complications and rely on the fact that even if the mentioned integration was to be performed, the temperature dependence
would still be governed to leading order by a Boltzmann weight $\exp(-\hbar\omega_0/k_BT)$.
We thus modify $\tau_\mathrm{mag}^{-1}$ and use 

\begin{equation}
 \tau_\mathrm{mag}^{-1}= C \theta(K-\hbar\omega)\frac{e^\frac{-\hbar\omega_0}{k_BT}}{1+3e^\frac{-\hbar\omega_0}{k_BT}}\label{iceblock_triplet}
\end{equation}
for fitting the data. As can be seen in Fig.~\ref{fig:comp_fit}, the quality of the corresponding fits is slightly improved with respect to the first resonant approach.

Finally, since one expects the excitation of a Kitaev system to fractionalise into Majorana fermions and gauge fluxes, one might speculate that the occupation function is governed by the statistics of fermions rather than that of triplets. We therefore have tested also whether a $\tau_\mathrm{mag}^{-1}$ of the form

\begin{equation}
 \tau_\mathrm{mag}^{-1}= C \theta(K-\hbar\omega)\frac{e^\frac{-\hbar\omega_0}{k_BT}}{1+e^\frac{-\hbar\omega_0}{k_BT}}\label{iceblock_fermions}
\end{equation}
improves the quality of the fit. However, in this case the fit quality is reduced, see Fig.~\ref{fig:comp_fit}. Thus, the further analysis of the data is performed using $\tau_\mathrm{mag}^{-1}$ in the form given in Eq.~\ref{iceblock_triplet}.

\subsection{Fitting procedure}
Upon fitting the data we assumed that the fitting parameters which describe the dynamics of the pure phononic system, i.e. which are captured in $\tau_{c,0}$ are independent of magnetic field. On the other hand we allow a field dependence  for $C$ and $\omega_0$. Furthermore, in the case of fitting with Eqs.~\ref{iceblock_triplet} and \ref{iceblock_fermions} we let  $K$ as a free parameter when fitting $\kappa_{ab}(T,B)$. When fitting $\kappa_{c}(T,B)$, the previously obtained value for $K$ is used as a fixed parameter.

To obtain good values for the field independent parameters describing conventional phonon scattering $\tau_{c,0}^{-1}$, multiple $\kappa (T)$ curves were fit simultaneously. 
Upon fixing the resulting phonon parameters, $C(B)$ and $w_0 (B)$ have then been optimised in a second step for each field separately. All fit parameters are given in the supplementary information in Tables~\ref{tab:field_indep} and \ref{tab:fielddep}. A comparison of the fits according to using a $\tau_\mathrm{mag}^{-1}$ as described by Eqs.~\ref{res_triplet}, \ref{iceblock_triplet}, and \ref{iceblock_fermions} for both $\kappa_{ab}$ and $\kappa_{c}$ are shown in Fig.~\ref{fig:comp_fit} in the supplementary information. After selecting Eq.~\ref{iceblock_triplet} as most successful for fitting the data, $C(B)$ and $\omega_0(B)$ have been determined for $\kappa_{ab}$ at all measured fields $B>12$~T, i.e. in steps of 0.5~T (the $\omega_0(B)$ are plotted as full squares in Fig.~\ref{fig:gap}). At smaller magnetic fields, these two parameters, and in particular $\omega_0$ cannot be determined very well because for those fields the measured $\kappa_{ab}(T)$ do not possess a clear enough minimum in the measured temperature range. Therefore, 
the extracted $\omega_0(B)$ for fields $B>12$~T are linearly extrapolated for estimating the field dependence of $\omega_0$ towards smaller fields. Thereby, an upper limit for $\omega_0(B)$ for $B\leq12$~T is be obtained (open squares in Fig.~\ref{fig:gap}). For those fields, fits to the data are obtained by fixing $\omega_0(B)$ to the extrapolated values and keeping only $C(B)$ as a free parameter separately for each field.   
Fig.~\ref{fig:abaxisfits} shows a direct comparison of the experimental data with these fits for selected magnetic fields. Furthermore, this figure contains a hypothetical curve where the magnetic scattering is switched of, i.e. $\tau_\mathrm{mag}^{-1}=0$.

For $\kappa_c$, the determination of $C(B)$ and $\omega_0(B)$ has been performed in steps of 1~T (see Table~\ref{tab:fielddep}).

\section{Acknowledgments}
This work has been supported by the Deutsche Forschungsgemeinschaft through SFB 1143 and through the projects HE3439/12 and HE3439/13. W.B. acknowledges partial support by QUANOMET, CiNNds, and PSM.
A.B. and S.E.N were supported by the Energy (US-DOE), Office of Science, Basic Energy Sciences (BES), Scientific User Facilities Division. 
D.G.M. and P.L.-K. were supported by the Gordon and Betty Moore Foundation's EPiQS Initiative through Grant GBMF4416.
J.S. and Y.-J.K. were supported by NSERC through Discovery Grant and CREATE program.

\section{Author contributions}
R.H. and A.U.B.W. carried out the heat transport and specific heat experiments, respectively. R.H., X.Z., W.B., and C.H. analysed the heat transport data. D.N., A.I., and T.D. grew and characterised samples I and III, P.L.-K., A.B., D.G.M., and S.E.N. grew and characterised sample II, J.S. and Y.-J.K. grew and characterised sample IV. B.B. and C.H. designed and supervised the project. R.H., W.B., and C.H. prepared the manuscript. All authors have read and approved the final version of the manuscript.

\section{Competing financial interests}
The authors declare no competing financial interests.

\bibliographystyle{naturemag_noURL}
\bibliography{Cuprates,rucl,frustrated_transport}

\newpage

\section{Supplementary Information}

\renewcommand{\thetable}{S\arabic{table}}

\renewcommand{\thefigure}{S\arabic{figure}}
\renewcommand{\thesubsection}{S\arabic{section}.\arabic{subsection}}
\renewcommand{\thesection}{S\arabic{section}}
\setcounter{section}{0}
\setcounter{page}{1}
\setcounter{figure}{0}
\setcounter{subsection}{0}

\subsection{Heat conductivity for different samples}

\subsection{In-plane heat transport}
\begin{figure*}[h!]
\centering
\includegraphics[width=\columnwidth]{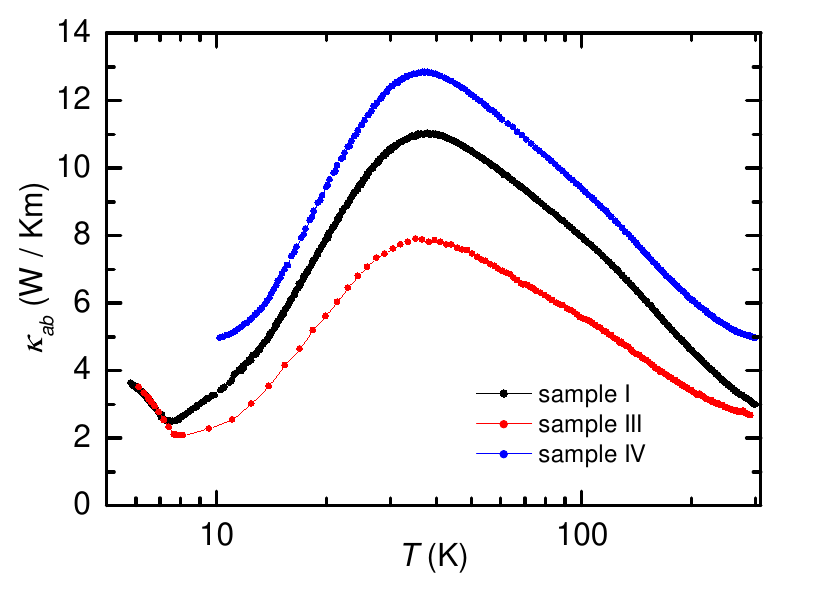}
\caption{In-plane thermal conductivity $\kappa _{ab}$ of samples I, III, and IV of $\alpha$-RuCl$_{3}$ single crystals (see main text for sample details). All samples surveyed show the same general features. }
\label{fig:threesamples}
\end{figure*}

Figure \ref{fig:threesamples} shows, additional to the $\kappa _{ab}(T)$ data at zero magnetic field of sample I discussed in the main text, the results of $\kappa _{ab}(T)$ measurements in zero field performed on samples III and IV. It is evident that the overall temperature dependence of all three samples exhibit the same features, a broad peak at $\sim \SI{40}{K}$ followed by a rapid decrease of $\kappa_{ab}$ towards lower temperatures and an eventual recovery below $T_N$ (temperature range not measured for sample IV). The peak position of $\kappa _{ab}$ displays only a minimal variation from sample to sample, which indicates that the sample purity with respect to point defect scatterers is practically the same. The nevertheless significant difference in absolute values can solely be explained by the uncertainty of the experimental geometry due to the small thickness and irregular sample shapes.

For $T>\SI{200}{K}$ a change of slope of $\kappa_ {ab} (T)$ becomes apparent which is more pronounced for sample IV than for sample I and III. 
One might conjecture, that the structural phase transition at 155~K \cite{Cao2016} where the honeycomb layers reorient with respect to each other plays a role in the change of slope of  $\kappa_{ab}(T)$. 
Furthermore,
at such high temperatures, unavoidable radiation losses generally become significant in heat transport experiments. These are sample dependent and thus are an additional possible explanation for the observed differences. 
We mention that for sample III, we measured $\kappa_\mathrm{ab}$ with a magnetic field parallel to the $c$-axis of the material. A large magnetic field dependence, as observed for fields parallel to the planes, has not been observed.

\subsection{Out-of-plane heat transport}
Upon measuring the out-of-plane heat conductivity on sample II, we observed hysteresis-like changes of $\kappa_c(T)$ near the structural phase transition at $T_S\sim \SI{155}{K}$ (not shown).
Therefore, the low-temperature measurements were performed after an initial cooling of the sample below $T_s$ without further crossing this transition.
Nevertheless, the out-of-plane heat conductivity of sample II displayed some instabilities during the measurements, which is indicative of the fragility of $\alpha$-RuCl$_{3}$'s crystal structure
reported earlier \cite{Cao2016}. We attribute this to unavoidable inhomogeneous thermal
strain during heating/cooling cycles which might change the initial stacking sequence along the $c$-direction, indubitably altering the phononic transport along the $c$-axis. Here, we discuss only the consecutively recorded, consistent part of the data. For $T>T_S$, we show the data recorded during heating the sample in zero field subsequent to the field dependent measurements.

\subsection{Field dependence of $\kappa_c$}

\begin{figure*}[h!]
\centering
\includegraphics[width=\columnwidth]{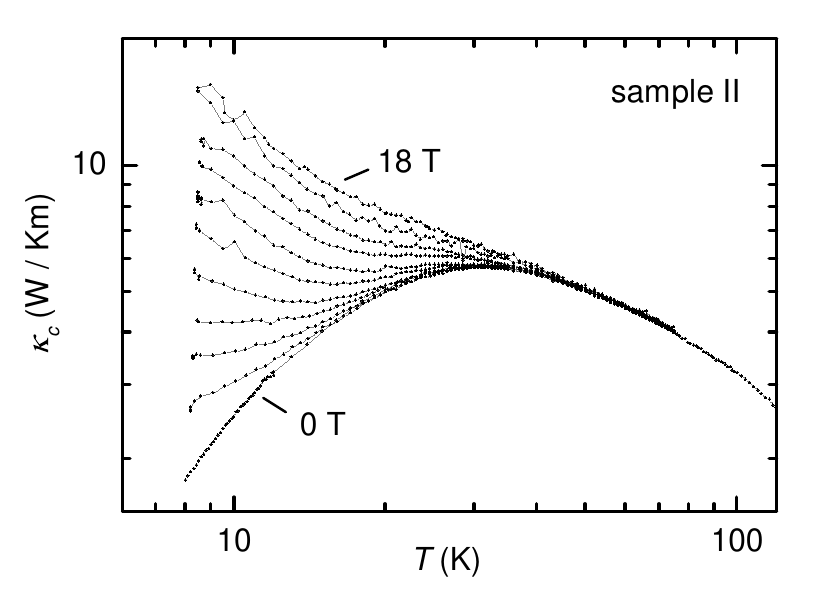}
\caption{Out-of-plane thermal conductivity $\kappa _{c}$ of $\alpha$-RuCl$_{3}$ measured on sample II with the magnetic fields of 0, 8, 10, 11, 12, 13, 14, 15, 16, 17, 18 Tesla applied parallel to the honeycomb planes.}
\label{fig:caxis}
\end{figure*}

\begin{figure*}[h!]
\centering
\includegraphics[width=\columnwidth]{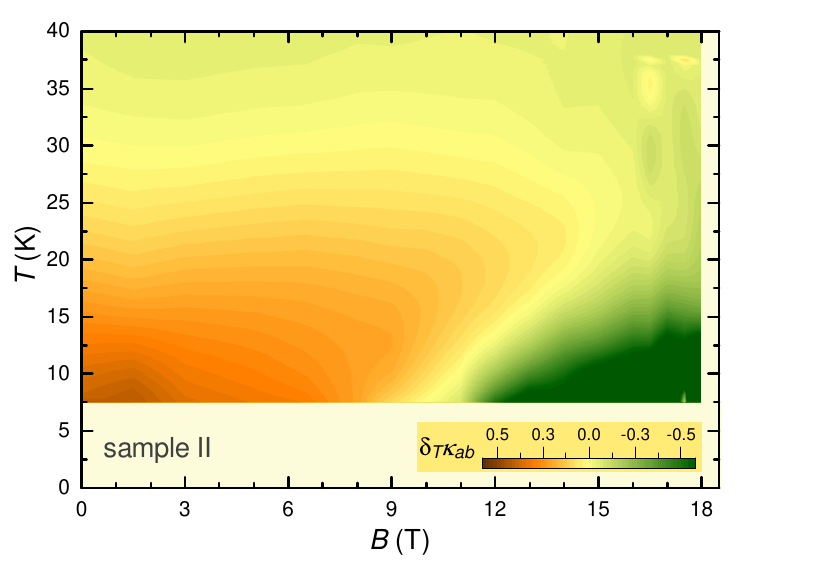}
\caption{False-colour representation of the temperature derivative $\partial\kappa_{c}(T, B)/\partial T$ (sample II). The approximately linear field dependence of the minimum in $\kappa_c$ for fields larger than about 10~T is clearly visible.}
\label{fig:caxisphd}
\end{figure*}

Fig.~\ref{fig:caxis} shows the low temperature heat conductivity measured on sample II, perpendicular to the honeycomb planes $\kappa _c$ at $B=\SI{0}{T}$ and selected $B>\SI{0}{}$. A low-temperature enhancement very similar to that of $\kappa_{ab}$ is clearly present. In order to track the magnetic field dependence of the minimum in $\kappa_c$, we plot in
Fig.~\ref{fig:caxisphd} the temperature derivative $\partial\kappa_c(T, B)/\partial T$ in false-colour representation. In analogy to the data for $\kappa_{ab}$ as shown in Fig.~\ref{fig:gap}, also here the minimum in $\kappa_c$ is clearly visible and approximately linear in $B$ for $B>B_c$.

\subsection{Fit results}

\begin{figure*}[h!]
\centering
\includegraphics[width=\textwidth]{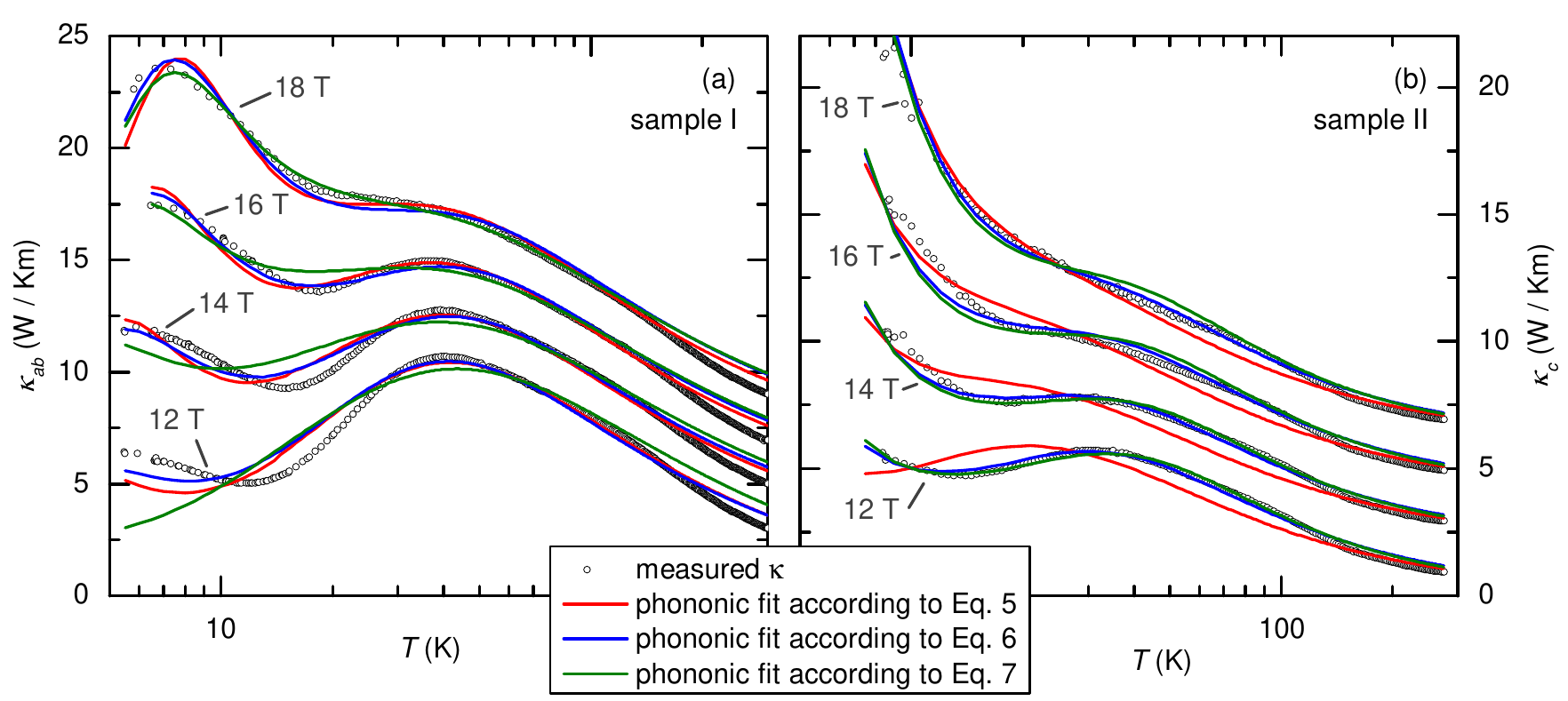}
\caption{$\kappa_{ab}$ (a) and $\kappa_{c}$ (b) data of samples I and sample II, respectively (open symbols) and fits to the Callaway model (solid lines) for selected magnetic fields. The fits have been obtained by incorporating magnetic scattering of the phonons using different mathematical forms for $\tau_\mathrm{mag}^{-1}$, according to  Eqs.~\ref{res_triplet},~\ref{iceblock_triplet}, and \ref{iceblock_fermions}  (see \textit{Methods}). }
\label{fig:comp_fit}
\end{figure*}

Fig.~\ref{fig:comp_fit} displays a comparison of the fit results according to the Callaway model (Eqs.~\ref{Callaway}, \ref{matthiessen} and \ref{comb}) under consideration of the different forms of $\tau_\mathrm{mag}^{-1}$ as described by Eqs.~\ref{res_triplet},~\ref{iceblock_triplet}, and \ref{iceblock_fermions} for both $\kappa_{ab}$ (sample I) and $\kappa_{c}$ (sample II). The field independent fit parameters for describing  $\kappa_{ab}$ and $\kappa_{c}$ with the Callaway model and $\tau_\mathrm{mag}^{-1}$ (Eq.~\ref{iceblock_triplet}) are given in Table~\ref{tab:field_indep}. The field dependent results for $C(B)$ and $\omega_0(B)$ for both  $\kappa_{ab}$  and $\kappa_{c}$ are reproduced in Table~\ref{tab:fielddep}.

\begin{figure*}[h!]
\centering
\includegraphics[width=\columnwidth]{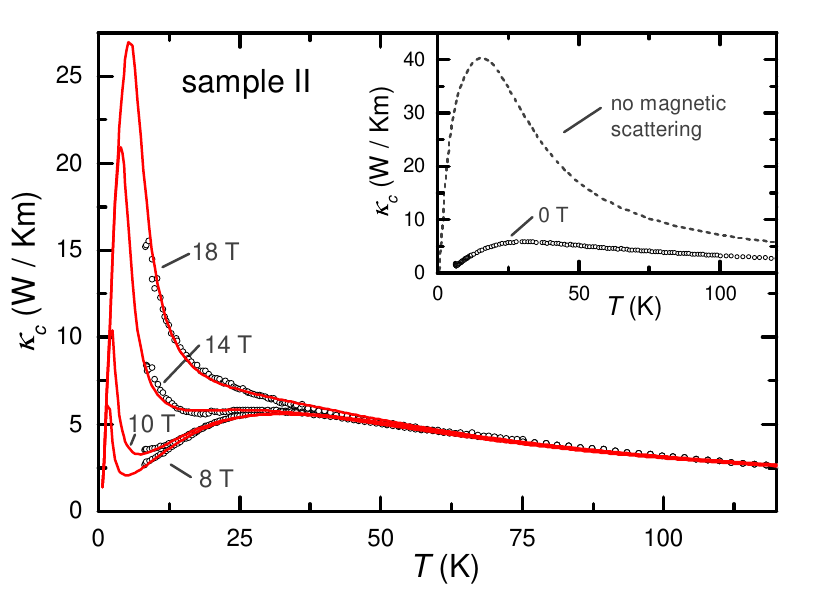}
\caption{$\kappa_{c}$ data of sample II (open symbols) and fits to the Callaway model (red solid lines) for  selected representative magnetic fields. The fits have been obtained by incorporating magnetic scattering of the phonons (see \textit{Methods}). Inset: $\kappa_{c}$ data of sample II (symbols) at zero field as compared to the hypothetical phononic heat conductivity described by the model if the magnetic scattering mechanism is switched off (solid line).}
\label{fig:caxis_fits}
\end{figure*}

Fig.~\ref{fig:caxis_fits} shows, in analogy to Fig.~\ref{fig:abaxisfits}, $\kappa_{c}$ data of sample II (open symbols) and fits to the Callaway model (red solid lines) for selected representative magnetic fields. 


\begin{table*}[!ht]
  
  \centering
  \begin{tabularx}{0.8\textwidth}{Y|Y|c|Y|Y|Y}

 \hspace{.2cm} Parameter \hspace{.2cm} &  $A$ (\SI{10^{-43}}{s^3}) &  $b$ &  $D$ (\SI{10^{-31}}{K^{-1}s^2}) & $L$ (\SI{10^{-2}}{m}) & $K/ k_B $(\SI{}{K}) \tabularnewline	 
\noalign{\smallskip} 	\hline \noalign{\smallskip}
	$\kappa_{ab}$ &  \SI{2.75}{}  & $\mspace{20mu} \SI{8.15}{}\mspace{20mu}$ & \SI{0.86}{} & \SI{0.13}{}& \SI{59.1}{}\tabularnewline
	 $\kappa_{c}$ &  \SI{11.18}{} &  \SI{3.22}{} & \SI{4.39}{} & \SI{2.14}{}& \SI{59.1}{} \tabularnewline \noalign{\smallskip} \hline
	 
	\end{tabularx}
	\caption{Field independent parameters for fitting $\kappa_{ab}$ and $\kappa_{c}$.}
  \label{tab:field_indep}
\end{table*}

\begin{table*}[!ht]
 
  \centering

    \begin{tabularx}{\textwidth}{Y|p{0.1\textwidth}|Y|Y|Y|Y|Y|Y|Y|Y|Y|Y|Y|Y}

\multicolumn{2}{c|}{Magnetic Field}     & \SI{12.5}{T}& \SI{13}{T}& \SI{13.5}{T}& \SI{14}{T}& \SI{14.5}{T}& \SI{15}{T}& \SI{15.5}{T}& \SI{16}{T}& \SI{16.5}{T}& \SI{17}{T}& \SI{17.5}{T}& \SI{18}{T}\tabularnewline	 
\noalign{\smallskip} 	\hline \noalign{\smallskip}
	\multicolumn{2}{c|}{$\kappa_{ab} \mspace{40mu} \hbar \omega _0 / k_B (\SI{}{K})\mspace{3mu}$}  & \SI{27.7}{}  &  \SI{30.1}{} & \SI{32.2}{} & \SI{34.3}{} & \SI{36.2}{} & \SI{38.2}{} & \SI{40.1}{} & \SI{42.1}{} & \SI{44.0}{} & \SI{45.9}{} & \SI{48.3}{} & \SI{49.9}{} \tabularnewline
	 \multicolumn{2}{c|}{ $\mspace{50mu} C (\SI{10^{8}}{s^{-1}})$}  & \SI{9.54}{}  &  \SI{9.77}{} & \SI{9.85}{} & \SI{9.93}{} & \SI{9.90}{} & \SI{9.89}{} & \SI{9.77}{} & \SI{9.68}{} & \SI{9.54}{} & \SI{9.44}{} & \SI{9.38}{} & \SI{9.16}{} \tabularnewline \noalign{\smallskip} \hline \noalign{\smallskip}
\multicolumn{2}{c|}{$\kappa_{c} \mspace{40mu} \hbar \omega _0 / k_B (\SI{}{K}) \mspace{3mu}$} & \SI {}{}  &  \SI{34.9}{} & \SI{ }{} & \SI{38.0}{} & \SI{ }{} & \SI{41.1}{} & \SI{ }{} & \SI{43.1}{} & \SI{ }{} & \SI{47.3}{} & \SI{}{} & \SI{54.4}{} \tabularnewline
	 \multicolumn{2}{c|}{ $\mspace{50mu} C (\SI{10^{8}}{s^{-1}})$} & \SI{ }{}  &  \SI{6.03}{} & \SI{ }{} & \SI{6.09}{} & \SI{ }{} & \SI{5.62}{} & \SI{ }{} & \SI{5.23}{} & \SI{ }{} & \SI{4.87}{} & \SI{ }{} & \SI{5.78}{} \tabularnewline \noalign{\smallskip} \hline
 
	\end{tabularx}

	\caption{Field dependent parameters for fitting $\kappa_{ab}$ and $\kappa_{c}$.}
  \label{tab:fielddep}
\end{table*}

\end{document}